\documentclass[article, aps, 12pt, floatfix, longbibliography, a4paper]{revtex4-2}
\usepackage{graphicx}
\usepackage{epstopdf}
\usepackage{subfigure}
\usepackage{hyperref}

\usepackage{color}
\usepackage[normalem]{ulem}

\usepackage[utf8]{inputenc}
\usepackage{xcolor}
\usepackage{physics}
\usepackage{textcomp}
\usepackage{multirow}

\renewcommand{\figurename}{\textbf{Fig.}}
\renewcommand{\tablename}{\textbf{Table}}
\makeatletter
\def\fnum@figure{\figurename\nobreakspace\textbf{\thefigure}}
\def\fnum@table{\tablename\nobreakspace\textbf{\thetable}}
\makeatother

\begin{document}

\title{Current-linear emergent induction of pinned skyrmion textures\\ in an oxide bilayer}

\author{Ludwig Scheuchenpflug$^{1}$,$^{\dagger}$}
\email{ludwig.scheuchenpflug@uni-a.de}
\author{Sebastian Esser$^{2}$}
\thanks{These two authors contributed equally.}
\author{Robert~Gruhl$^{1}$}
\author{Max Hirschberger$^{2,3}$}
\author{Philipp Gegenwart$^{1}$}\email{philipp.gegenwart@physik.uni-augsburg.de}

\affiliation{$^{1}$Experimentalphysik VI, Center for Electronic Correlations and Magnetism, University of Augsburg, D-86159 Augsburg, Germany}
\affiliation{$^{2}$Department of Applied Physics and Quantum-Phase Electronics Center (QPEC), The University of Tokyo, Bunkyo-ku, Tokyo 113-8656, Japan}
\affiliation{$^{3}$RIKEN Center for Emergent Matter Science (CEMS), Wako, Saitama, 351-0198, Japan}
\maketitle

\newpage
\begin{center}
\Large{Abstract}
\end{center}
Emergent electromagnetic induction (EEMI) induced through current-driven spin dynamics was recently predicted and subsequently observed in helical spin magnets, opening a new direction in spintronics and paving the way towards further miniaturization of electronic circuit elements. In contrast to conventional inductors consisting of coil-like structures whose inductance $L$ shows a linear dependence on the cross-section $A$, emergent inductors exhibit an inverse ($\propto {A}^{-1}$) proportionality, favorable for the miniaturization of electrical devices. However, the expected current-linear response of the EEMI voltage has not been demonstrated. Magnetic skyrmions hold promise as a simple platform to study the conceptual foundations of EEMI from current-driven spin dynamics. We fabricated devices of thin film bilayers of ferromagnetic SrRuO$_3$ and paramagnetic SrIrO$_3$, which are known to host interfacial N\'eel skyrmions detected by the appearance of a topological Hall-effect (THE). A large, positive and current-linear inductive response is found to accompany the THE. In our experiment, the current-induced dynamics of pinned magnetic skyrmions creates a voltage both parallel and perpendicular to the applied electric current flow, corresponding to longitudinal and transverse induction, respectively. This is the first report of transverse EEMI, indicating an angle of $80^{\circ}$ between skyrmion motion and the applied current. Our observation of current-linear longitudinal and transverse EEMI is a hallmark of pinned dynamics of magnetic skyrmion textures in oxide heterostructures.

\begin{center}
\Large{Introduction}
\end{center}
Sustaining Moore's law for electronic miniaturization is a major challenge for near-future developments in conventional circuitry and should comprise all the standard components of electronic circuits, including inductive reactance. Inductors are currently based on coil-like structures in which, when an alternating current $I$ is applied, the induced voltage is given by $U=L~\mathrm{d}I/\mathrm{d}t$. The inductance $L$ is proportional to the cross-section $A$ of the magnetic core \cite{Jackson1998}, severely limiting further miniaturization steps. A solution of this challenging problem, an inductor with a predicted inverse proportionality to the cross-section $L\sim 1/A$, was proposed recently. The prediction is based on a new quantum mechanical concept for current-driven spin dynamics in spin-helix systems \cite{Nagaosa2019}, termed emergent electromagnetic induction. The theory describes time-dependent spin dynamics, which generates a Berry vector potential $\mathbf{a}$ \cite{Berry1984, Xiao2010} and its associated emergent magnetic ($B^\mathrm{em}_i$) and electric ($E^\mathrm{em}_i$) fields \cite{Nagaosa2019}
\begin{align}
    B^\mathrm{em}_i &= \left( \nabla \times \mathbf{a} \right)_i = \frac{h}{8\pi e} \varepsilon_{ijk} \mathbf{n} \cdot \left( \partial_j \mathbf{n} \times \partial_k \mathbf{n} \right) \label{eq:bem} \\
    E^\mathrm{em}_i &= -\frac{1}{c} \partial_t \mathbf{a} = \frac{h}{2\pi e} \mathbf{n} \cdot \left( \partial_i \mathbf{n} \times \partial_t \mathbf{n} \right) \label{eq:EEMI}
\end{align}
where $\mathbf{n}$ is a unit vector along the direction of the local spin moment, $\varepsilon_{ijk}$ is the total asymmetric tensor and Einstein convention is adopted. The time-dependent deformation of non-collinear spin structures due to spin-transfer torque is described by $\mathbf{n}(t)$, driven by an alternating current (AC)~\cite{Xiao2010, Nagaosa2013}, and leads to a time-varying magnetic moment along the helical axis \cite{Nagaosa2019, Yokouchi2020}. The induced electric field can then be detected via the spin-motive force~\cite{Ohe2009,Tanabe2012}, as described by Eq.~(\ref{eq:EEMI}).

In addition to further theoretical work~\cite{Kurebayashi2021, Ieda2021} and micromagnetic simulations~\cite{Furuta2023, Furuta2023b}, first experimental realizations of emergent inductance were proposed in microstructured, short-pitch helimagnets~\cite{Yokouchi2020, Kitaori2021, Kitaori2024, Kitaori2023} as well as in permalloy thin film ferromagnets with domain walls~\cite{Matsushima2024}. However, the linearity of the induced voltage $U$ in the applied current density $j$, corresponding to $j$-independence of $L$, has never been demonstrated~\cite{Yokouchi2020, Kitaori2024, Kitaori2021}. This lack of a current-linear response is not compatible with the theoretical description and obstructs technical application. 

\begin{figure}[h]
\centering
\includegraphics[width=0.98\textwidth]{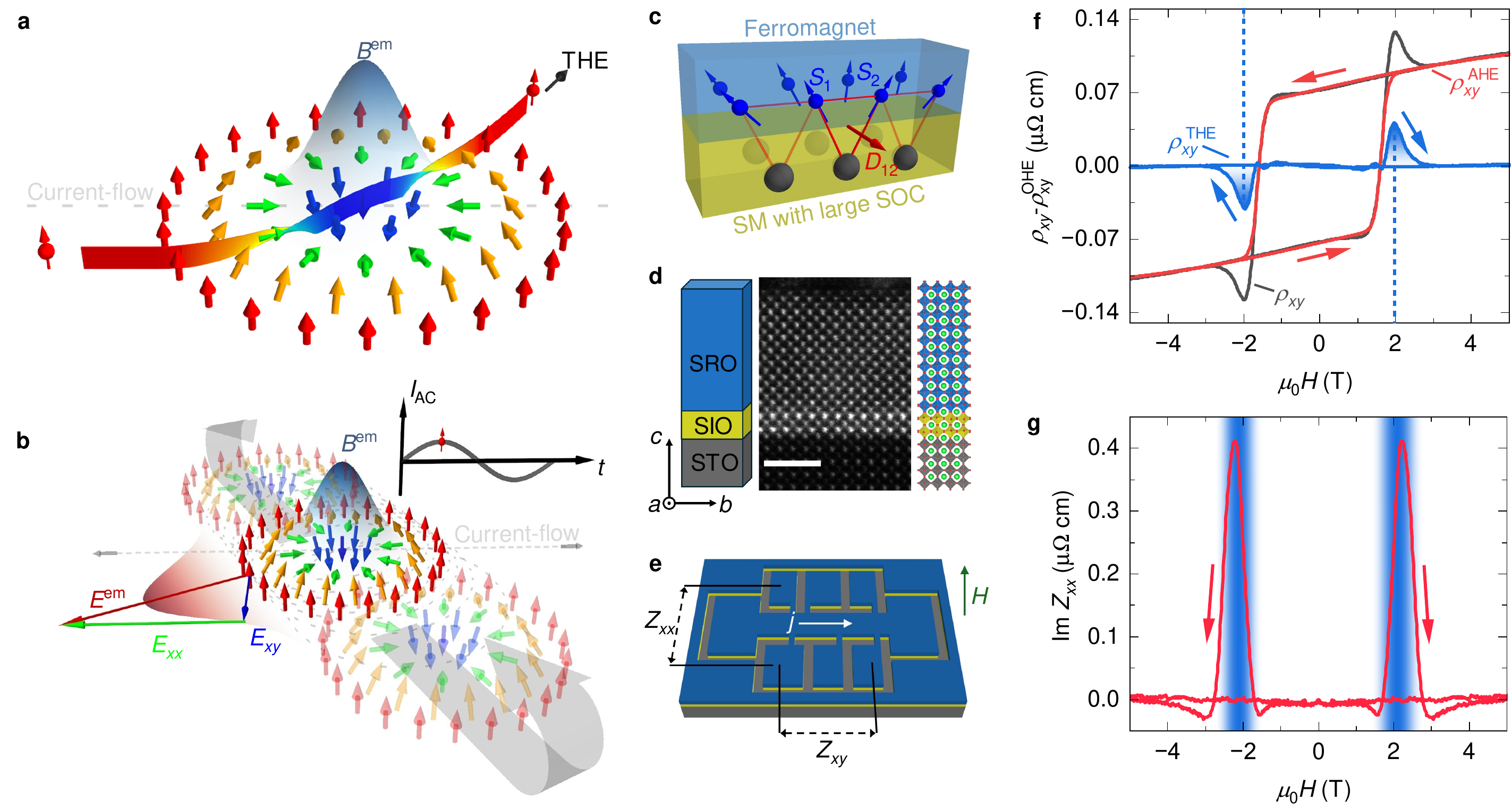}
\caption{\textbf{Concept of EEMI from pinned interface skyrmions in the bilayer oxide system SrRuO$_\mathbf{3}$/SrIrO$_\mathbf{3}$.}  \textbf{a} Schematic: Interplay of a skyrmion and conduction electrons, resulting in the topological Hall effect, due to the emergent magnetic field $B^{\mathrm{em}}$ occurring by adiabatic alignment of the conduction electron spin to the magnetic structure. \textbf{b} Schematic: Representation of the emergent induction of a pinned skyrmion in an external alternating current. The emergent electric field $E^{\mathrm{em}}$ is generated by the AC current-driven oscillating motion of the skyrmion and the associated time-dependent change in the emergent magnetic field $B^{\mathrm{em}}$. \textbf{c} Non-vanishing Dzyaloshinskii-Moriya interaction $D_{12}$ due to the proximity of a semi-metal (SM) with large spin-orbit coupling (SOC) to a ferromagnetic metal as driving force behind the skyrmion formation in SrRuO$_3$ (SRO) / SrIrO$_3$ (SIO) bilayers \cite{Matsuno2016, Esser21}. \textbf{d} High-resolution TEM image of the SRO-SIO bilayer structure grown on SrTiO$_3$ (STO), with $n=2\,\mathrm{uc}$ and $m=10\,\mathrm{uc}$ (unit cells) for SIO and SRO respectively. The white scale-bar corresponds to $2\,\mathrm{nm}$ and the high atomic number $Z$-contrast of the elements indicates sharp interfaces. \textbf{e} Hall-bar like geometry prepared by the focused ion beam (FIB) technique for the measurement of the components of the complex impedance tensor $Z_{ij}$. \textbf{f} DC Hall resistance $\rho_{xy}$ after subtraction of the contribution from the ordinary Hall effect (OHE). Following Ref. \cite{Esser21} the contribution of the topological Hall effect (THE, blue curve) was extracted from by a two-channel model fit that also accounts for the anomalous Hall effect (red line). \textbf{g} Magnetic field dependence of the longitudinal imaginary impedance ($\mathrm{Im} Z_{xx}$) measured with $j = 10^{8}\,\mathrm{A}\,\mathrm{m}^{-2}$, $f = 7.5\,\mathrm{kHz}$ and $\mathbf{H}$ in out-of-plane direction. The blue shaded area corresponds to regions with non-vanishing THE in panel f.}\label{fig1}
\end{figure}

In this work, we investigate the EEMI of pinned magnetic skyrmions \cite{Nagaosa2013}, which was predicted from numerical model calculations in Ref. \cite{Furuta2023b}. Compared to a spin helix, the skyrmions considered here have -- due to their topological charge -- a static magnetic field $B^{\mathrm{em}}$, calculated according to Eq.~(\ref{eq:bem}), which leads to a deflection of electrons passing through the spin texture. This phenomenon causes the topological Hall effect (THE) as shown in Fig.~\ref{fig1}~\textbf{a}. Accordingly, due to $B^{\mathrm{em}}$ and the current-driven low-energy dynamics of skyrmions, an emergent electric field $E^{\mathrm{em}}$ is expected in the form of an induction. We report a large longitudinal and a small transversal EEMI, which correspond to voltages parallel and perpendicular to the applied current density $\mathbf{j}$, respectively. Contrary to prior reports of current-nonlinear, negative inductance in helical magnets~\cite{Yokouchi2020,Kitaori2021,Kitaori2023,Kitaori2024}, the longitudinal EEMI in our skyrmion hosting bilayer has a positive sign and satisfies the requirements of linear response theory.

Thin films are advantageous for investigations of EEMI, as the measurement geometry can be precisely controlled by designing suitable devices. Most importantly for EEMI studies, the cross-section $A$ of the film can much smaller than the one of mesoscopic devices cut from bulk samples. This allows for higher current densities $j=I/A$ despite lower total current $I$. In addition, Joule heating generated by $I$ is readily dissipated into the substrate due to the large surface-to-volume ratio of the bilayer film~\cite{Yokouchi_rebuttal}. 

\begin{center}
\Large{Results}
\end{center}

\textbf{EEMI of magnetic skyrmions: mechanism}\\
The coupling between skyrmions and an applied current results not only in the THE but also in oscillatory motion of skyrmions, if the current density is not too high. The dynamics of skyrmions in response to a current is described by the Landau-Lifshitz-Gilbert-Slonczewski equation \cite{SLONCZEWSKI1996L1, Shibata2006}, which implies that the skyrmions move at an angle of $\alpha$ to the applied charge current -- the skyrmion Hall effect. Due to the current-induced motion of the skyrmions, their $\mathbf{B}^\mathrm{em}$ changes as a function of time, leading to an emergent electric field $\mathbf{E}^\mathrm{em}$ analogous to Faraday's law of induction, described by Eq.~(\ref{eq:EEMI}). Note, that in this case $\mathbf{E}^\mathrm{em}$ can also be rewritten as \cite{Zang2011, Schulz2012}:
\begin{align}
    \mathbf{E}^\mathrm{em} = \mathbf{v}_\mathrm{sk} \times \mathbf{B}^{em}\mathrm{,}
    \label{eq:vsk}
\end{align}
where $\mathbf{v}_\mathrm{sk}$ is the velocity of the skyrmions. Skyrmions moving with the applied current $\mathbf{j}$ enter the continuous flow regime at current densities of $j \sim 10^{10} \,\mathrm{Am^{-2}}$ \cite{Jiang_2017}. Here, the induced voltage resulting from $\mathbf{E}^\mathrm{em}$ is in-phase with the applied current and opposes the Hall voltage stemming from $\mathbf{B}^\mathrm{em}$. This leads to a reduction of the observed THE~\cite{Zang2011, Schulz2012} or even its suppression to zero when the skyrmion and conduction electrons move in the same reference frame; this latter limit is associated with emergent Galilean invariance~\cite{Birch2024}. In our work, $j$ is much lower, on the order of $10^{5}\,\mathrm{Am}^{-2}$ to $10^{8}\,\mathrm{Am}^{-2}$, significantly below the critical current density for skyrmion motion. Thus, we are investigating skyrmion dynamics at low excitation energies, where skyrmions are trapped at pinning centers and cannot move over wide distances, but rather oscillate around a potential energy minimum following the applied alternating current, as illustrated in Fig.~\ref{fig1}~\textbf{b}. Neglecting dissipation, the position of the skyrmion follows the applied sinusoidal current flow and is farthest from the pinning center when the current is highest. At the instant when no current is flowing, the skyrmion is at the pinning center but reaches its maximum velocity. At this instance, the change in the emergent magnetic field is greatest, causing an extremum of the emergent electric field. The speed of the skyrmion is thus proportional to the derivative of the current $v_\mathrm{Sk} \propto \dd I/\dd t$ and, using Eq.~(\ref{eq:vsk}), the voltage defined by $\mathbf{E}^{\mathrm{em}}$ appears out of phase to the applied current, in the form of an inductance.\\

\textbf{Hall effect and EEMI in SRO-SIO bilayers}\\
To experimentally investigate the inductance resulting from skyrmion dynamics, perovskite metal oxides represent a suitable material platform due to their high tunability and crystalline quality. Breaking the inversion symmetry at an atomically-sharp interface between the strongly spin-orbit coupled semimetal SrIrO$_3$ (SIO) and the itinerant ferromagnet SrRuO$_3$ (SRO), we introduce a Dzyaloshinskii-Moriya interaction (DMI) which is known to be a driving force for skyrmion formation \cite{Fert2013}, cf. Fig.\ref{fig1}~\textbf{c}. N\'eel-skyrmions in such bilayers were first reported based on the appearance of a topological Hall effect (THE) \cite{Matsuno2016, Esser21} and later confirmed by precise imaging experiments~\cite{Meng2019}, reporting a small skyrmion radius of $r_{\mathrm{Sk}} \sim 10-20\,\mathrm{nm}$. These rather small skyrmions are advantageous for the investigation of the EEMI effect, since short-period spin textures generate a larger inductive response: For example, the EEMI of a spin helix grows with the inverse of its helical pitch~\cite{Nagaosa2019, Yokouchi2020}. 
Epitaxial bilayers -- consisting of 2-10 unit cells (uc) SIO and $10\,$uc SRO -- were grown on $(001)$-oriented SrTiO$_3$ (STO) substrates as shown in Fig.~\ref{fig1}~\textbf{d}. Their high structural quality was confirmed by X-ray diffraction and transmission electron microscopy, details of which are depicted in \ref{sec: S4}. Microstructures for electrical transport measurements with different cross-sections and electrode spacings, cf. Fig.~\ref{fig1}~\textbf{e} and \ref{Supfig5}, were created using the focused ion beam (FIB) technique. 

To confirm the existence of skyrmions at the interface, the Hall resistivity $\rho_{xy}$ in the DC-limit was measured first, where three contributions to $\rho_{xy}$ have been taken into account, cf. Fig.~\ref{fig1}~\textbf{f}: the ordinary $\rho_{xy}^{\mathrm{OHE}}$, the anomalous $\rho_{xy}^{\mathrm{AHE}}$ and the topological $\rho_{xy}^{\mathrm{THE}}$ Hall resistivity. While $\rho_{xy}^{\mathrm{OHE}} \sim H$ and $\rho_{xy}^{\mathrm{AHE}} \sim M(H)$, $\rho^{\mathrm{THE}}_{xy} = R_0PB^{\mathrm{em}}$ \cite{Neubauer2009} depends on the strength of the emergent magnetic field $B^{\mathrm{em}}$. Here $R_0$ is the Hall coefficient and $P$ is the effective spin polarization. The additional topological contribution to the Hall resistivity is shown in Fig.~\ref{fig1}~\textbf{f} (blue curve), which is extracted by removing $\rho_{xy}^{\mathrm{OHE}}$ and $\rho_{xy}^{\mathrm{AHE}}$ as in Ref.~\cite{Esser21}.

For measurements of the complex impedance, an AC current drive was applied to the sample in a standard lock-in technique. Figure~\ref{fig1}~\textbf{g} presents an example of the background-subtracted imaginary part of the impedance $\mathrm{Im} Z_{xx}$ measured at a temperature of $T=2\,\mathrm{K}$, a frequency of $f = 7.5\,\mathrm{kHz}$ and with a current density of $j = 10^{8} \,\mathrm{Am^{-2}}$. Two distinct maxima in $\mathrm{Im} Z_{xx}$ emerge in the same field range as the THE (blue shaded area), while no such signal appears in the saturated ferromagnetic state. This observation clearly links the EEMI to the appearance of magnetic skyrmions \cite{Meng2019}. 
Reproducibility of the inductive signal was ensured by measuring several bilayers in different measurement geometries, as listed in Supplementary Table~\ref{tabS1}. The detailed frequency dependence of the EEMI is discussed in \ref{sec: fabh}, including limitations of the experiment at $>10 \,$kHz due to spurious capacitive couplings.\\

\textbf{Symmetry of the EEMI tensor}\\
Ref.~\cite{Furuta2023b} shows that the emergent inductance can be represented as a second-order tensor $\mathbf{L}$, which also includes a transverse component $L_{xy}$. Furthermore, $\mathbf{L}$ must be a symmetric tensor in the adiabatic limit of the Landau-Lifshitz-Gilbert-Slonczewski formalism, a condition that is derived on energetic and symmetry grounds and confirmed by micromagnetic simulations~\cite{Furuta2023b}.

\begin{figure}[h]
\centering
\includegraphics[width=0.98\textwidth]{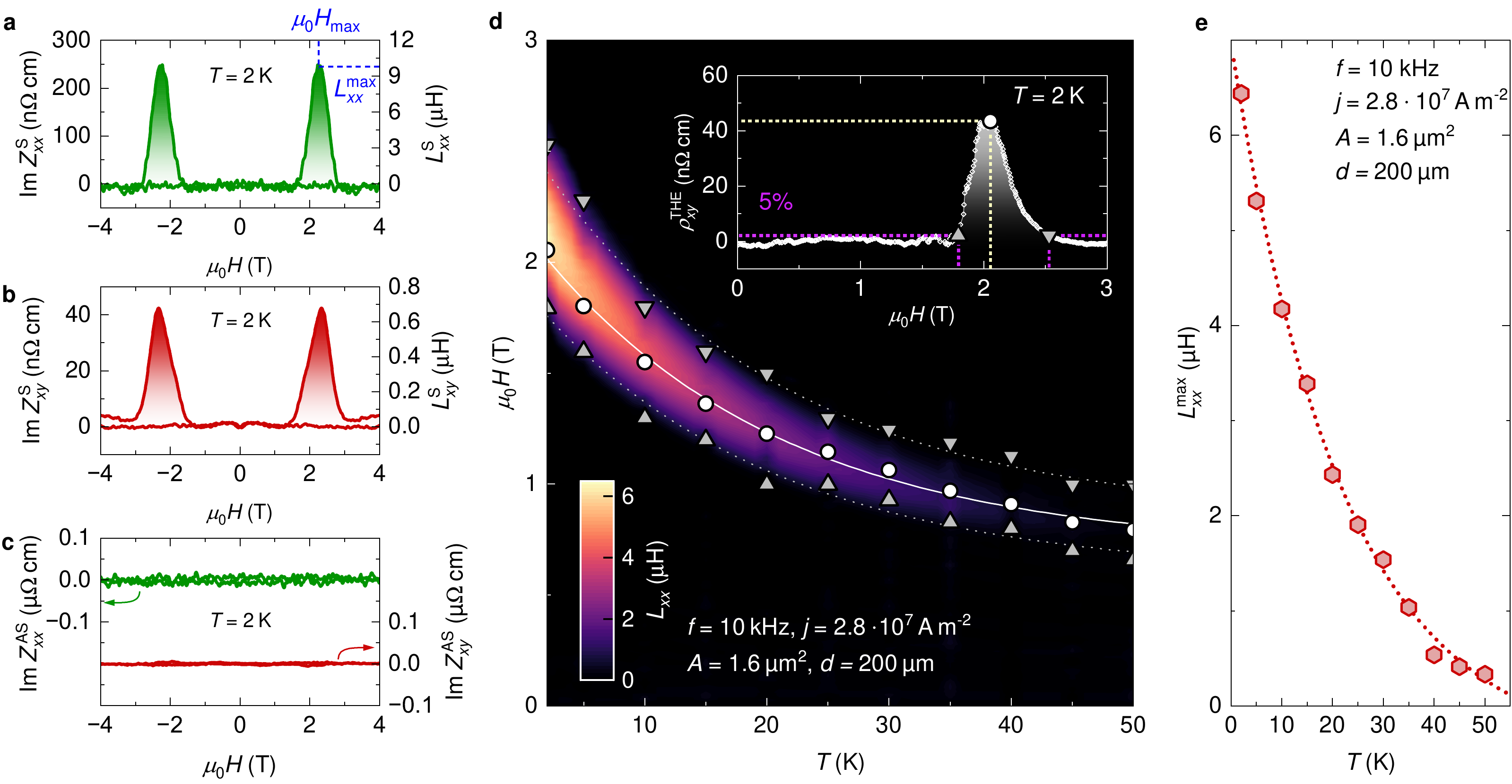}
\caption{\textbf{Symmetry of the EEMI tensor and its temperature dependence.} \textbf{a} - \textbf{c} Magnetic field dependence of the field-symmetric (a,b) and -antisymmetric (c) imaginary parts of the complex impedance tensor $Z_{ij}$ measured with $j = 3.8\cdot 10^{7}\,\mathrm{A}\,\mathrm{m}^{-2}$, $f = 10\,\mathrm{kHz}$, $A = 2$\,\textmu$\mathrm{m}^2$, $d = 500$\,\textmu$\mathrm{m}$ and $\mathbf{H}$ perpendicular to the surface of the device. The transverse inductance indicates a skyrmion movement that is slightly tilted from the expected perpendicular movement to the current flow. \textbf{d} Temperature and magnetic field dependence of the inductance $L_{xx}$ together with the center position of the topological Hall effect (white circles) and its occurrence area (grey triangles). Lines are a guide to the eye. Inset: Topological Hall effect for the same sample. Peak position (yellow line) and threshold ($5\,\%$ of peak value, purple lines) are used to determine the region of interest in panel d. \textbf{e} Temperature dependence of the maximum value of the inductance $L^{\mathrm{max}}_{xx}$ extracted from the field sweeps in panel d.}\label{fig2}
\end{figure}

For a skyrmion lattice or individual skyrmions in the continuum limit, the pinned, adiabatic motion of the magnetic vortex in linear response to a current $\mathbf{j}$ has a velocity $\mathbf{v_\mathrm{Sk}}$ that is perfectly perpendicular to $\mathbf{j}$. The resulting emergent electric field is described by Eq.~(\ref{eq:vsk}), is parallel to $\mathbf{j}$, and is a symmetric function of the external magnetic field. This high-symmetry scenario thus produces a diagonal inductance tensor of the type $\mathbf{L} = L_{ij}\delta_{ij}$ without off-diagonal elements~\cite{Furuta2023}. Experimentally, however, we observe a small, off-diagonal $L_{xy}^{\mathrm{S}}$ element which is symmetric to the external magnetic field $\mathbf{H}$. Therefore, our experiment indicates a breaking of rotation symmetry in the environment of the skyrmion, e.g. by coupling to the underlying orthorhombic crystal lattice of the  perovskite SRO-SIO bilayer. In Fig.~\ref{fig2}~\textbf{a}~-~\textbf{c} we show the imaginary impedance signals after symmetrization to the magnetic field: The longitudinal $\mathrm{Im} Z^{\mathrm{S}}_{xx}$ (\textbf{a}) and transverse $\mathrm{Im} Z^{\mathrm{S}}_{xy}$ (\textbf{b}) as well as their antisymmetric counterparts $\mathrm{Im} Z^{\mathrm{A}}_{xx}$ and $\mathrm{Im} Z^{\mathrm{A}}_{xy}$ (\textbf{c}). While the antisymmetric impedances are nearly zero, the symmetric longitudinal impedance is significantly larger than the symmetric transverse impedance. The labels $S$ and $A$ are therefore omitted in the following. From $\mathrm{Im}Z_{ij} = 2\pi fL_{ij}$ \cite{Jackson1998} the inductance can be calculated from these symmetrized imaginary impedances. Our experiment represents the first observation of an off-diagonal element of the emergent inductance tensor. In the bilayers, $\mathbf{v_\mathrm{Sk}}$ is not perfectly perpendicular to $\mathbf{j}$; instead, a slightly tilted motion ($\alpha \sim 80^\circ$) is reported, cf. \ref{sec: sk_motion}.\\

\textbf{Rapid decay of EEMI due to thermal fluctuations}\\
To further investigate the EEMI of skyrmions at the SRO-SIO oxide interface, the temper\-ature-dependent imaginary impedance $\mathrm{Im}Z_{xx}$ was measured with a frequency of $f = 10\,\mathrm{kHz}$ and a current density of $j = 2.8 \cdot 10^{7}\,\mathrm{Am^{-2}}$, shown in Fig.~\ref{fig2}~\textbf{d}. The inductive signal appears only in the temperature and field range of finite topological Hall effect, as measured in the DC limit. This further supports the notion that both signals can be attributed to skyrmions. White circles in Fig.~\ref{fig2}~\textbf{d} depict the peak position of $\rho_{xy}^{\mathrm{THE}}$ and gray triangles the field values where the Hall signal drops below 5\% of its peak value, cf. inset of panel \textbf{d}. These symbols outline the field range where the skyrmion-driven THE is observed in our oxide bilayer. The inductance reaches values of around $L^{\mathrm{max}}_{xx} = 7.5$\,\textmu$\mathrm{H}$ at a temperature of $2\,$K with a sample cross-section of $A = 1.6$\,\textmu$\mathrm{m}^2$ and an electrode distance of $d = 200$\,\textmu$\mathrm{m}$. Increasing the temperature, the $L^{\mathrm{max}}_{xx}$ declines rapidly and vanishes above $50\,\mathrm{K}$, cf. Fig.~\ref{fig2}~\textbf{e}. The much more rapid decay of $L^{\mathrm{max}}_{xx}$ as compared to $\rho_{xy}^{\mathrm{THE, max}}$, cf. \ref{sec: S2}, indicates an important role for temperature-induced damping in the skyrmions dynamics. \\

\begin{figure}[h]
\centering
\includegraphics[width=0.98\textwidth]{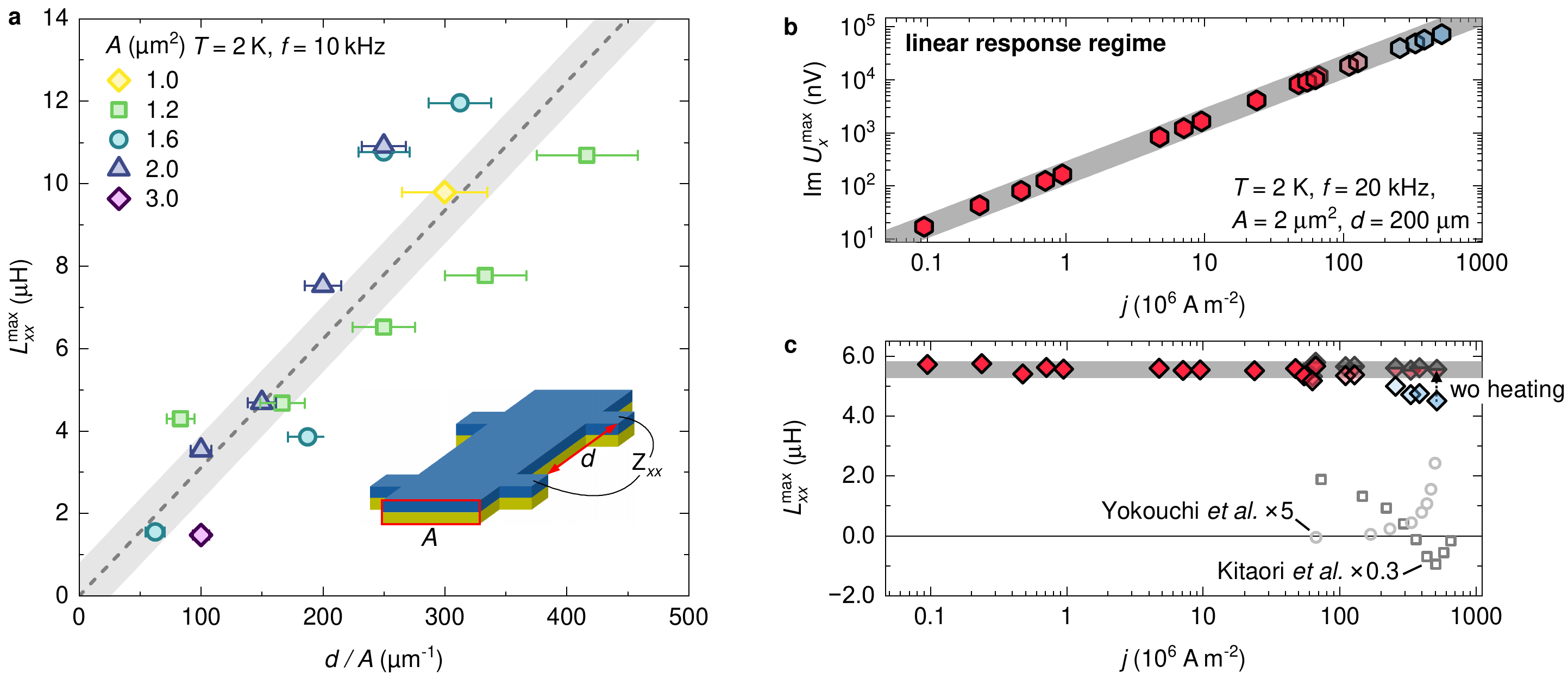}
\caption{\textbf{Current-linear inductive response in bilayer SrRuO$_\mathbf{3}$/SrIrO$_\mathbf{3}$ devices.} \textbf{a} Scaling plot of the inductance $L^{\mathrm{max}}_{xx}$ with electrode spacing $d$ and cross-section $A$ for different SrRuO$_3$/SrIrO$_3$ bilayer devices as listed in Supplementary Table~\ref{tabS1}. \textbf{b} Current density $j$ dependence of the longitudinal voltage drop $\mathrm{Im}\,U^{\mathrm{max}}_{x}$, measured for a device with $A = 2$\,\textmu$\mathrm{m}^{2}$ and $d = 200$\,\textmu$\mathrm{m}$ at $f = 20\,\mathrm{kHz}$ and $T = 2\,\mathrm{K}$. The red filled dots show a linear response to the applied current density over several orders of magnitude in $j$. \textbf{c} Inductance associated with this voltage, $L_{xx}^\mathrm{max}$, which is independent of $j$ within this range of $j$. Deviations from the linear response behavior for current densities above $j \sim 5\cdot10^7\,\mathrm{Am^{-2}}$ (blue symbols in both panels) can be explained by heating effects, cf. \ref{sec: S1}. Quantitatively taking heating into account, the reduction of $L_{xx}^\mathrm{max}$ can be numerically compensated (half filled symbols) using the temperature dependence of $L_{xx}^\mathrm{max}$, cf. Fig.~\ref{fig2}~\textbf{e} and Supplementary~Fig.~\ref{Supfig_Iabh_2}~\textbf{c}. For comparison, current-dependent literature data of EEMI (gray circles and squares) from \cite{Yokouchi2020,Kitaori2024} are shown, which are not linear in $j$.}\label{fig3}
\end{figure}

\textbf{EEMI in the linear response regime}\\
Figure~\ref{fig3}~\textbf{a} shows the geometry dependence of the inductance for a series of bilayer devices, where the inductance $L^{\mathrm{max}}_{xx}$ is plotted against the ratio $d/A$ of contact distance $d$ and device cross-sectional area $A$. In the simple theory of Ref.~\cite{Nagaosa2019}, $L^{\mathrm{max}}_{xx}$ is linear in $d/A$. While the agreement of theory and experiment is rather good, our experimental data do not exactly follow the expected $L\propto {d}/{A}$ behavior. Deviations can originate from inactive regions of the sample, e.g. at the outer edge of the device, which may be created during the microstructuring process.

All available theoretical models for EEMI consider the regime of linear response, where the applied current serves as a probe of low-energy spin dynamics in helical magnets~\cite{Nagaosa2019, Kurebayashi2021, Ieda2021}. In our skyrmion hosting bilayers, the voltage drop $\mathrm{Im}U^{\mathrm{max}}_{x}$ at $T = 2\,\mathrm{K}$ and $f = 20\,$kHz follows this theoretical prediction of a current-linear response over several orders of magnitude in $j$, cf. Fig.~\ref{fig3}~\textbf{b}. The corresponding inductance shown in Fig.~\ref{fig3}~\textbf{c} is therefore constant in this range, which contrasts with prior work on EEMI in helical-spin magnets, where the linear response regime could not be realized and $L$ typically depends on the square of the applied current density~\cite{Yokouchi2020, Kitaori2021,Kitaori2024}. We indicate this previous work by gray circles and squares in Fig.~\ref{fig3}~\textbf{c}. 

In our experiment, the deviation from $\mathrm{Im}U^{\mathrm{max}}_{x}\propto j$ at current densities above \linebreak $j\sim5\cdot10^{7}\,\mathrm{Am^{-2}}$ is caused by Joule heating; the temperature increase leads to a decrease in $L_{xx}$ as depicted in Fig.~\ref{fig2}~\textbf{e}. The device temperature can be monitored and $L^{\mathrm{max}}_{xx}$ can be corrected for the heating effect -- details are shown in  \ref{sec: S1}.

The half-filled symbols in Fig.~\ref{fig3}~\textbf{c} show $L_{xx}^\mathrm{max}$ after the correction. The inductance remains independent of $j$ for another decade in $j$, consistent with linear response theory. The observed current dependence also confirms that the skyrmions in SRO-SIO remain in the pinned regime over the entire range of experimentally probed current densities. A much sharper change of inductance and reactance would be expected close to and above the depinning threshold~\cite{Birch2024}.\\

\textbf{Perspective}\\
We report a positive emergent electromagnetic inductance (EEMI) consistent with linear response theory in bilayers of the spin-orbit coupled semimetal SrIrO$_3$ with the ferromagnetic metal SrRuO$_3$. Our discovery offers various opportunities in fundamental research and technology: First, extended frequency and field-angle studies can provide deep insight into the dynamics of small interfacial skyrmions. This type of pinned, low-energy dynamics of topological magnetic solitons is difficult to access by established methods. We have also experimentally demonstrated transverse induction, which corresponds to an angle $\alpha\neq 90^\circ$. As the off-diagonal component of the inductance tensor represents a kind of mutual inductance and is highly sensitive to the local symmetry of magnetic textures~\cite{Furuta2023b}, this opens new possibilities for inductive components controllable via strain or chemical composition tuning. Our observations on iridate-ruthenate bilayer devices illustrate oxide heterostructures as a suitable and broad material class for realizing EEMI in the current-linear regime, as required for applications. Demonstrating this type of EEMI at room temperature in devices with sufficiently high $T_{\mathrm{C}}$ values can 
pave the way for the utilization of EEMI in next-generation technology.

\clearpage
\begin{center}
\Large{Methods}
\end{center}

\textbf{Sample preparation and characterization}\\
Bilayers of SrIrO$_3$ ($n=2\,\mathrm{or}\,10\,$unit\,cells, uc) and SrRuO$_3$ ($m=10\,\mathrm{uc}$), cf. Fig.~\ref{fig1}~\textbf{d}, were grown by the metalorganic aerosol deposition technique~\cite{Jungbauer2014} on $(001)$ oriented SrTiO$_3$ substrates. Crystal quality, phase purity, strain state and the thickness of the thin films were measured in a Malvern Panalytical Empyrean diffractometer operating with Cu-K$\alpha$ radiation and with a hybrid monochromator. The X-ray reflectometry pattern was simulated using the GenX software package~\cite{GenX}. High-resolution scanning transmission electron microscopy (HR-STEM) studies were carried out using a  JEOL JEM-ARM200F NEOARM transmission electron microscope (TEM) equipped with a Cs-corrector. The HR-STEM images were taken in cross-section geometry with a high-angle annular dark-field (HAADF) detector and an acceleration voltage of $200\,\mathrm{keV}$. Exemplary TEM images are shown in Supplementary Fig.~\ref{Supfig3}. The TEM lamellae were prepared, lifted out, thinned and cleaned using a Zeiss Crossbeam 550 focused ion beam (FIB) equipped with an up to $30\, \mathrm{keV}$ Ga ion-beam and an additional scanning electron microscope (SEM). \\

\textbf{Electrical transport measurements}\\
For the electrical transport measurements, each bilayer was microstructured into Hall bar-like structures with the help of the dual beam FIB, cf. Fig.~\ref{fig1}~\textbf{e} and Supplementary Fig.~\ref{Supfig5}, with different cross-sections $A$ and electrode spacings $d$, as listed in Supplementary Table~\ref{tabS1}. The temperature and magnetic field dependences of the DC electrical resistivity $\rho_{xx}$ and the Hall resistivity $\rho_{xy}$, as well as the frequency dependent components of the complex impedance $Z_{ij}$, were determined in a Physical Properties Measurement System (PPMS) equipped with a $14\,\mathrm{T}$ magnet and a Dynacool system with a $9\,\mathrm{T}$ magnet from Quantum Design (USA). Voltage drops across the device were recorded by DSP-SR830 Lock-in amplifiers from Stanford Research Systems and digital MFLI Lock-in amplifiers from Z{\"u}rich Instruments.
An adjustable AC voltage $U=U_0\sin{(\omega t)}$ was applied across a precise shunt resistance ($R_\mathrm{p} = 50\,\mathrm{k}\Omega$) in series with the structured sample ($R_\mathrm{s}\sim 2\,\mathrm{k}\Omega$). Thus, the injected AC current $I=I_0\sin{(\omega t)}$ is fixed. In our four-terminal geometry, we recorded in-phase $U_{j,X}^{1f}$ and out-of-phase $U_{j,Y}^{1f}$ components of the first harmonic voltage simultaneously, for both resistance and Hall contacts. The complex impedance was calculated as 
\begin{align}
    Z_{ij}^{1f} = \frac{U_{j,X}^{1f}}{I_0}+\imath \frac{U_{j,Y}^{1f}}{I_0}.
\end{align}
The applied current density $j = I_0/A$ was varied over a broad range from $10^{5}\,\mathrm{Am^{-2}}$ to $5\cdot10^{8}\,\mathrm{Am^{-2}}$, corresponding to an applied current $I_0 = 50\,\mathrm{nA}$ to $0.25\,\mathrm{mA}$. When measuring the complex impedance, $Z_{xx}^{1f}$ and $Z_{xy}^{1f}$ can contribute to the measured voltage drop at the same contact pair, making it necessary to apply symmetry operations to separate the transverse and longitudinal impedance $Z_{xy}^{1f}$ and $Z_{xx}^{1f}$, respectively. Therefore, the measured signal was symmetrized to obtain $Z_{xx}^{1f}$ and antisymmetrized to obtain $Z_{xy}^{1f}$. As the inductive response in $\mathrm{Im}Z_{xy}^{1f}$ must be symmetric, this contribution was also symmetrized. The frequency dependence of the longitudinal inductance was determined by performing several measurements at fixed temperature and current, varying the frequency. On this basis, further measurements were carried out in the frequency-independent range  -- details in \ref{sec: fabh}. 

Before each field ramp for each individual temperature, frequency or current dependent measurement, the out-of-phase component of $Z$ at maximum magnetic field was minimized by adjusting the phase of the injected current with respect to the measured voltage drop. Other parasitic impedances, such as those caused by the microstructure or by the wiring of the measurement setup, lead to additional phase shifts. Background subtraction is required to obtain the reported impedance signal. Further information can be found in the \ref{sec: S3}.\\

\begin{center}
\Large{Data availability}
\end{center}
The data supporting the findings of this study are available from the authors upon reasonable request.

\clearpage
\begin{center}
\Large{References}
\end{center}
\bibliographystyle{naturemag}

\begin{thebibliography}{31}
\providecommand{\url}[1]{\texttt{#1}}
\expandafter\ifx\csname urlstyle\endcsname\relax
  \providecommand{\doi}[1]{doi: #1}\else
  \providecommand{\doi}{doi: \begingroup \urlstyle{rm}\Url}\fi

\bibitem[1]{Jackson1998}
{Jackson}, J.~D.
\newblock Classical Electrodynamics.
\newblock \emph{John Wiley} \& \emph{Sons} (1998).

\bibitem[2]{Nagaosa2019}
{Nagaosa}, N.
\newblock Emergent inductor by spiral magnets.
\newblock \emph{Japanese Journal of Applied Physics} \textbf{58}, 120909 (2019).

\bibitem[3]{Berry1984}
{Berry}, M.~V.
\newblock Quantal phase factors accompanying adiabatic changes.
\newblock \emph{Proceedings of the Royal Society of London A} \textbf{392}, 45--57 (1984).

\bibitem[4]{Xiao2010}
{Xiao}, D., {Chang}, M.-C. and {Niu}, Q.
\newblock Berry phase effects on electronic properties.
\newblock \emph{Reviews of Modern Physics} \textbf{82}, 1959--2007 (2010).

\bibitem[5]{Nagaosa2013}
{Nagaosa}, N. and {Tokura}, Y.
\newblock Topological properties and dynamics of magnetic skyrmions.
\newblock \emph{Nature Nanotechnology} \textbf{8}, 899--911 (2013).

\bibitem[6]{Yokouchi2020}
{Yokouchi}, T., {Kagawa}, F., Hirschberger, {M.} \textit{et al.}
\newblock Emergent electromagnetic induction in a helical-spin magnet.
\newblock \emph{Nature} \textbf{586}, 232-236 (2020).

\bibitem[7]{Ohe2009}
{Ohe}, J. and {Maekawa}, S.
\newblock Spin motive force in magnetic nanostructures.
\newblock \emph{Journal of Applied Physics} 105, 07C706 (2009).

\bibitem[8]{Tanabe2012}
{Tanabe}, K., {Chiba}, D., {Ohe}, J. \textit{et al.}
\newblock Spin-motive force due to a gyrating magnetic vortex.
\newblock \emph{Nature Communications} \textbf{3}, 845 (2012).

\bibitem[9]{Kurebayashi2021}
{Kurebayashi}, D. and {Nagaosa}, N.:
\newblock Electromagnetic response in spiral magnets and emergent inductance.
\newblock \emph{Communications Physics} \textbf{4} (2021).

\bibitem[10]{Ieda2021}
{Ieda}, J. and {Yamane}, Y.
\newblock Intrinsic and extrinsic tunability of Rashba spin-orbit coupled
  emergent inductors.
\newblock \emph{Physical Review B} \textbf{103}, L100402 (2021).

\bibitem[11]{Furuta2023}
{Furuta}, S., {Moody}, S.~H., {Kado}, K. \textit{et al.}
\newblock Energetic perspective on emergent inductance exhibited by magnetic
  textures in the pinned regime.
\newblock \emph{npj Spintronics} \textbf{1}, 1 (2023).

\bibitem[12]{Furuta2023b}
{Furuta}, S., {Koshibae}, W. and {Kagawa}, F. 
\newblock Symmetry of the emergent inductance tensor exhibited by magnetic
  textures.
\newblock \emph{npj Spintronics} \textbf{1}, 3 (2023).

\bibitem[13]{Kitaori2021}
{Kitaori}, A., {Kanazawa}, N., {Yokouchi}, T. \textit{et al.}
\newblock Emergent electromagnetic induction beyond room temperature.
\newblock \emph{Proceedings of the National Academy of Sciences} \textbf{118}, e2105422118
  (2021).

\bibitem[14]{Kitaori2024}
{Kitaori}, A., {White}, J.~S., {Ukleev}, V. \textit{et al.}
\newblock Enhanced emergent electromagnetic inductance in
  $\mathrm{Tb}_5\mathrm{Sb}_3$ due to highly disordered helimagnetism.
\newblock \emph{Communications Physics} \textbf{7}, 159 (2024).

\bibitem[15]{Kitaori2023}
{Kitaori}, A., {White}, J.~S., {Kanazawa}, N. \textit{et al.}
\newblock Doping control of magnetism and emergent electromagnetic induction in
  high-temperature helimagnets.
\newblock \emph{Physical Review B} \textbf{107}, 024406 (2023).

\bibitem[16]{Matsushima2024}
{Matsushima}, Y., {Zhang}, Z., {Ohashi}, Y. \textit{et al.}
\newblock Emergent magneto-inductance effect in permalloy thin films on
  flexible polycarbonate substrates at room temperature.
\newblock \emph{Applied Physics Letters} \textbf{124}, 022404 (2024).

\bibitem[17]{Yokouchi_rebuttal}
{Yokouchi}, T., {Kitaori}, A., {Yamaguchi}, D. \textit{et al.}
\newblock Comment on "Reconsidering the nonlinear emergent inductance:
  time-varying Joule heating and its impact on the AC electrical response".
\newblock  	Preprint at \url{https://arxiv.org/abs/2407.15682} (2024).

\bibitem[18]{SLONCZEWSKI1996L1}
{Slonczewski}, J.C.
\newblock Current-driven excitation of magnetic multilayers.
\newblock \emph{Journal of Magnetism and Magnetic Materials} \textbf{159}, L1-L7 (1996).

\bibitem[19]{Shibata2006}
{Shibata}, J., {Nakatani}, Y., {Tatara}, G. \textit{et al.}
\newblock Current-induced magnetic vortex motion by spin-transfer torque.
\newblock \emph{Physical Review B} \textbf{73}, 020403 (2006).

\bibitem[20]{Zang2011}
{Zang}, J., {Mostovoy}, M., {Han}, J.~H. and {Nagaosa}, N.
\newblock Dynamics of Skyrmion Crystals in Metallic Thin Films.
\newblock \emph{Physical Review Letters} \textbf{107}, 136804 (2011).

\bibitem[21]{Schulz2012}
{Schulz}, T., {Ritz}, R. and {Bauer}, A.
\newblock Emergent electrodynamics of skyrmions in a chiral magnet.
\newblock \emph{Nature Physics} \textbf{8}, 301--304 (2012).

\bibitem[22]{Jiang_2017}
{Jiang}, W., {Zhang}, X., {Yu}, G. \textit{et al.}
\newblock Direct observation of the skyrmion Hall effect.
\newblock\emph{Nature Physics} \textbf{13},  162--169 (2017).

\bibitem[23]{Birch2024}
{Birch}, M.~T., {Belopolski}, I., {Fujishiro}, Y. \textit{et al.}
\newblock Dynamic transition and Galilean relativity of current-driven
  skyrmions.
\newblock \emph{Nature} \textbf{633}, 554--559 (2024).

\bibitem[24]{Fert2013}
{Fert}, A., {Cros}, V. and {Sampaio}, J.
\newblock Skyrmions on the track.
\newblock \emph{Nature Nanotechnology} \textbf{8}, 152--156 (2013).

\bibitem[25]{Matsuno2016}
{Matsuno}, J., {Ogawa}, N., {Yasuda}, K. \textit{et al.}
\newblock Interface-driven topological Hall effect in
  $\mathrm{SrRuO}_3$-$\mathrm{SrIrO}_3$ bilayer.
\newblock \emph{Science Advances} \textbf{2}, e1600304 (2016).

\bibitem[26]{Esser21}
{Esser}, S., {Wu}, J., {Esser}, S. \textit{et al.}
\newblock Angular dependence of Hall effect and magnetoresistance in
  $\mathrm{SrRuO}_{3}$-$\mathrm{SrIrO}_{3}$ heterostructures.
\newblock \emph{Physical Review B} \textbf{103}, 214430 (2021).

\bibitem[27]{Meng2019}
{Meng}, K.-Y., {Ahmed}, A., {Ba{\'c}ani}, M. \textit{et al.}
\newblock Observation of {Nanoscale} {Skyrmions} in {SrIrO$_3$}/{SrRuO$_3$}
  {Bilayers}.
\newblock \emph{Nano Letters} \textbf{19}, 3169--3175 (2019).

\bibitem[28]{Neubauer2009}
{Neubauer}, A., {Pfleiderer}, C., {Binz}, B. \textit{et al.}
\newblock Topological {Hall} {Effect} in the {A} {Phase} of {MnSi}.
\newblock \emph{Physical Review Letters} \textbf{102}, 186602 (2009).

\bibitem[29]{Jungbauer2014}
{Jungbauer}, M., {H{\"u}hn}, S., {Egoavil}, R. \textit{et al.}
\newblock Atomic layer epitaxy of Ruddlesden-Popper SrO(SrTiO$_3$)$_n$ films by
  means of metalorganic aerosol deposition.
\newblock \emph{Applied Physics Letters} \textbf{105}, 251603 (2014).

\bibitem[30]{GenX}
{Bj{\"{o}}rck}, M. and {Andersson}, G.
\newblock {{GenX}: an extensible X-ray reflectivity refinement program
  utilizing differential evolution}.
\newblock \emph{Journal of Applied Crystallography} \textbf{40}, 1174--1178 (2007).
\end{thebibliography}

\begin{center}
\Large{Acknowledgments}
\end{center}
This work was funded by the Deutsche Forschungsgemeinschaft (DFG, German Research Foundation) through TRR 360-49258816. M.H. was supported by the Japan Science and Technology Agency (JST) as part of Adopting Sustainable Partnerships for Innovative Research Ecosystem (ASPIRE), Grant Number JPMJAP2426, and by JST FOREST (JPMJFR2238). S.E. and M.H. also acknowledge support from the Japan Society for the Promotion of Science (JSPS) under Grant Nos. JP22F22742, JP23H05431 and JP24H01607.

\begin{center}
\Large{Author contributions}
\end{center}
P.G. and S.E. conceived the project. L.S. and R.G. synthesized and characterized the thin films. R.G. prepared the microstructured devices and performed the TEM analysis. Electric transport measurements were carried out by L.S. and were analyzed under the guidance of M.H. and S.E.. All authors wrote the manuscript.

\begin{center}
\Large{Competing interests}
\end{center}
The authors declare no competing interests.

\clearpage
\begin{center}
\Large{Main Text Figures}
\end{center}
\enlargethispage{10\baselineskip}

\clearpage
\renewcommand\thefigure{S\arabic{figure}} 
\setcounter{figure}{0}

\renewcommand\thesubsection{Supplementary Note~\arabic{subsection}} 
\setcounter{section}{0}   

\renewcommand\thetable{S\arabic{table}} 
\setcounter{table}{0}   

\begin{center}
\Large{Supplementary Material}
\end{center}
\subsection{Current dependent Im$Z$-measurements}\label{sec: S1}
\begin{figure}[!b]
\centering
\includegraphics[width=0.9\textwidth]{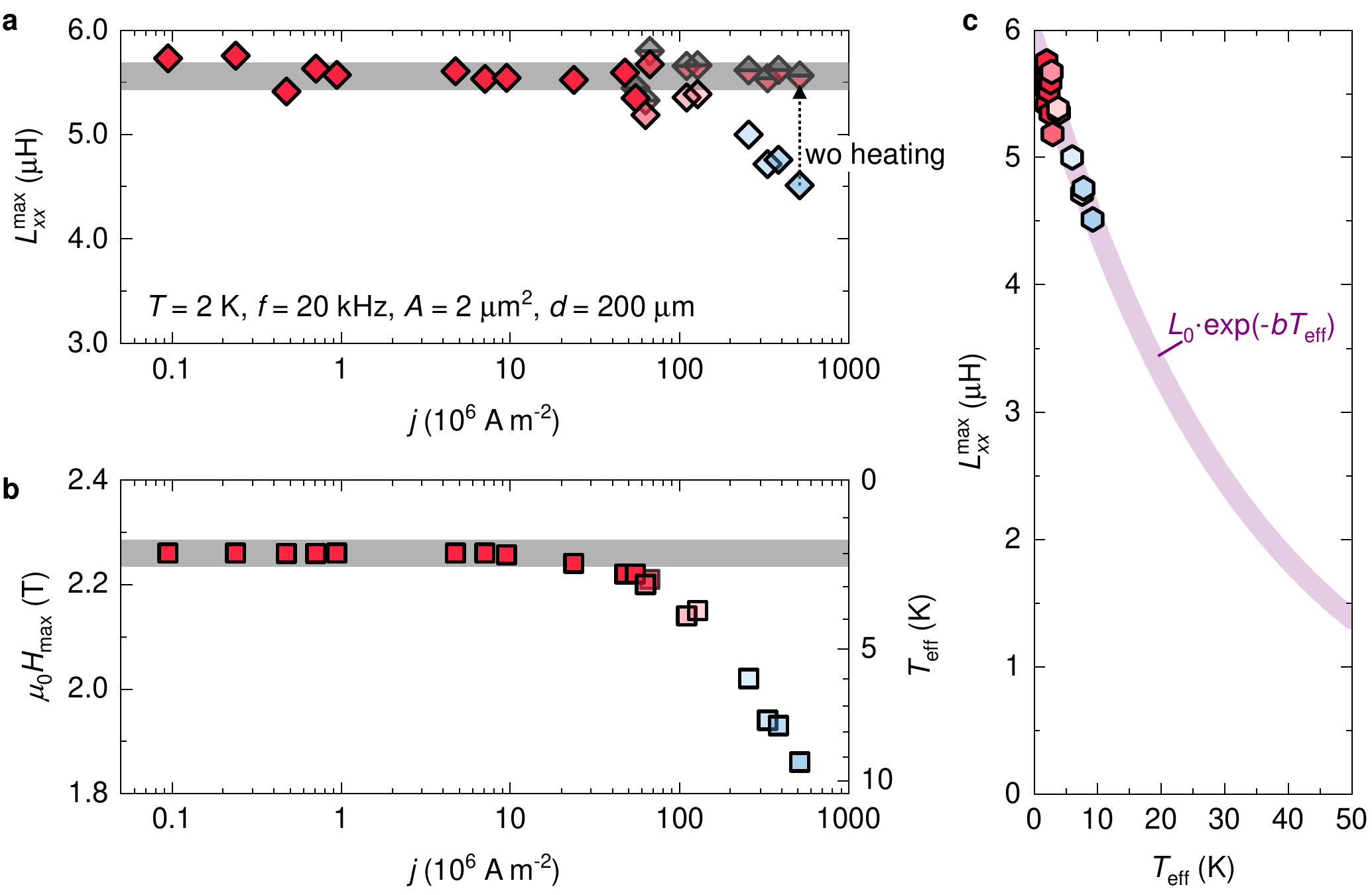}\caption{\textbf{Decline of EEMI-signal due to Joule heating.}
 \textbf{a} Current density $j$ dependence of the inductance $L^{\mathrm{max}}_{xx}$ measured for a device with $A = 2$\,\textmu$\mathrm{m}^{2}$ and $d = 200$\,\textmu$\mathrm{m}$ at $f =20\,\mathrm{kHz}$ and a heat bath temperature of $T=2\,\mathrm{K}$. For current densities higher then $\sim5\cdot 10^7\,\mathrm{Am^{-2}}$ the inductive signal declines as a result of Joule heating. \textbf{b} Shift of the peak position $\mu_0H_{\mathrm{max}}$ of $L^{\mathrm{max}}_{xx}$ due to Joule heating. The effective temperature $T_\mathrm{eff}$ (right axis) was calculated from the temperature and magnetic field dependent $L$, cf. Main Text Fig.~\ref{fig2}~\textbf{d}. \textbf{c} Effective, temperature dependent decay of the longitudinal inductance $L^{\mathrm{max}}_{xx}$ of this sample based on data from panel b. The decay was fitted using an exponential form (purple line) with $L_0 = 5.93$\,\textmu$\mathrm{H}$ and $b = 0.029\,\mathrm{K}^{-1}$.}\label{Supfig_Iabh_2}
\end{figure}
Figure~\ref{Supfig_Iabh_2}~\textbf{a} shows the current density $j$ dependence of $L^{\mathrm{max}}_{xx}$ as in Fig.~\ref{fig3}~\textbf{b} of the Main Text, which is constant over several orders of magnitude in $j$. At $j\sim5\cdot 10^7\,\mathrm{Am^{-2}}$ the induction starts to decrease slightly (filled symbols), which is attributed to heating effects. Considering the shift of the peak position $\mu_0H_{\mathrm{max}}$ from $L^{\mathrm{max}}_{xx}$ with increasing current density, cf. Fig.~\ref{Supfig_Iabh_2}~\textbf{b}, an effective sample temperature $T_\mathrm{eff}$ can be calculated from the $T$-$H$ diagram of $L$ --  Main Text Fig.~\ref{fig2}~\textbf{d}. This $T_\mathrm{eff}$ is shown on the right axis of Fig.~\ref{Supfig_Iabh_2}~\textbf{b}. An increase of $T_\mathrm{eff}$ leads to a decrease of $L_{xx}^\mathrm{max}$ as depicted in Fig.~\ref{fig2}~\textbf{e} of the Main Text.
Modeling the temperature decay with a phenomenological exponential form, we extract the coefficient $b = 0.029\,\mathrm{K}^{-1}$ for this sample, c.f. purple line in Fig.~\ref{Supfig_Iabh_2}~\textbf{c}. Thus, we calculate the corrected longitudinal inductance at $T=2\,\mathrm{K}$ at each current density
\begin{align}
    L_\mathrm{xx}^{max} \left(T=2\,\mathrm{K}\right) = L_\mathrm{xx}^{max}\left(\mathrm{measured~at}~T_\mathrm{eff}\right)\cdot\exp\left[-b(T_\mathrm{eff}-T)\right].
\end{align}
The resulting values of $L_\mathrm{xx}^{max}$ are depicted in Fig.~\ref{Supfig_Iabh_2}~\textbf{a} by half filled symbols. After the correction, $L_{xx}^\mathrm{max}$ is independent of the current density $j$ (grey line) -- i.e., consistent with linear response -- over another order of magnitude. The successful correction of our EEMI data for the Joule heating effect in the device confirms that the EEMI originates from the magnetic properties of the device itself, not from the circuit. It further emphasizes or confirms the presence of Joule heating at higher current densities.

\begin{figure}[!bh]
\centering
\includegraphics[width=0.573\textwidth]{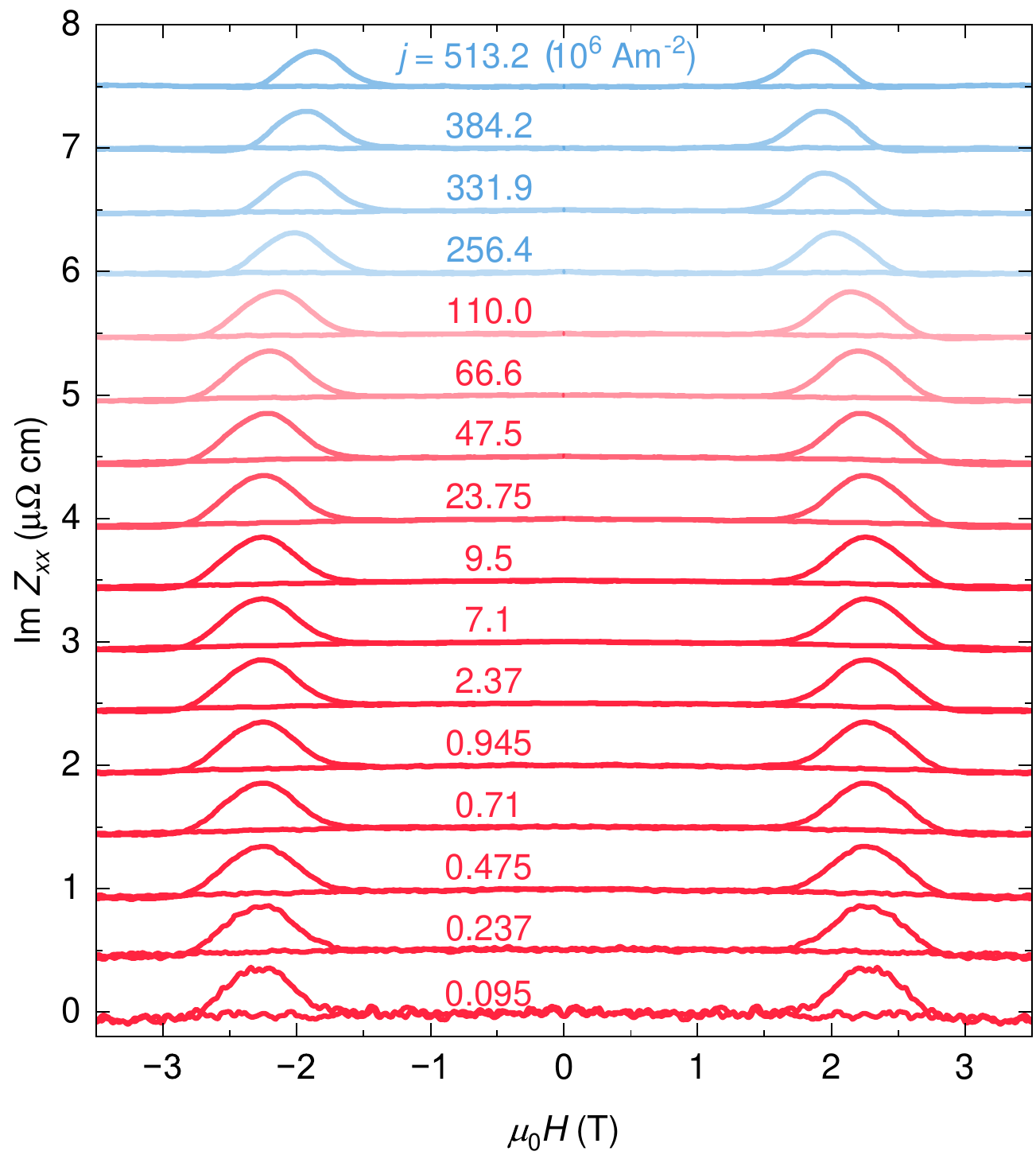}
\caption{\textbf{Current density dependence of the imaginary part of the longitudinal impedance.} Magnetic field dependent imaginary impedance for various current densities $j$ at $T = 2\,\mathrm{K}$ with $f = 20\,\mathrm{kHz}$, $d = 200$\,\textmu$\mathrm{m}$ and $A = 2.0$\,\textmu$\mathrm{m ^{2}}$.}\label{Supfig1}
\end{figure}

\clearpage
\subsection{Temperature dependent Im$Z$-measurements and THE}\label{sec: S2}
\begin{figure}[b]
\centering
\includegraphics[width=0.6\textwidth]{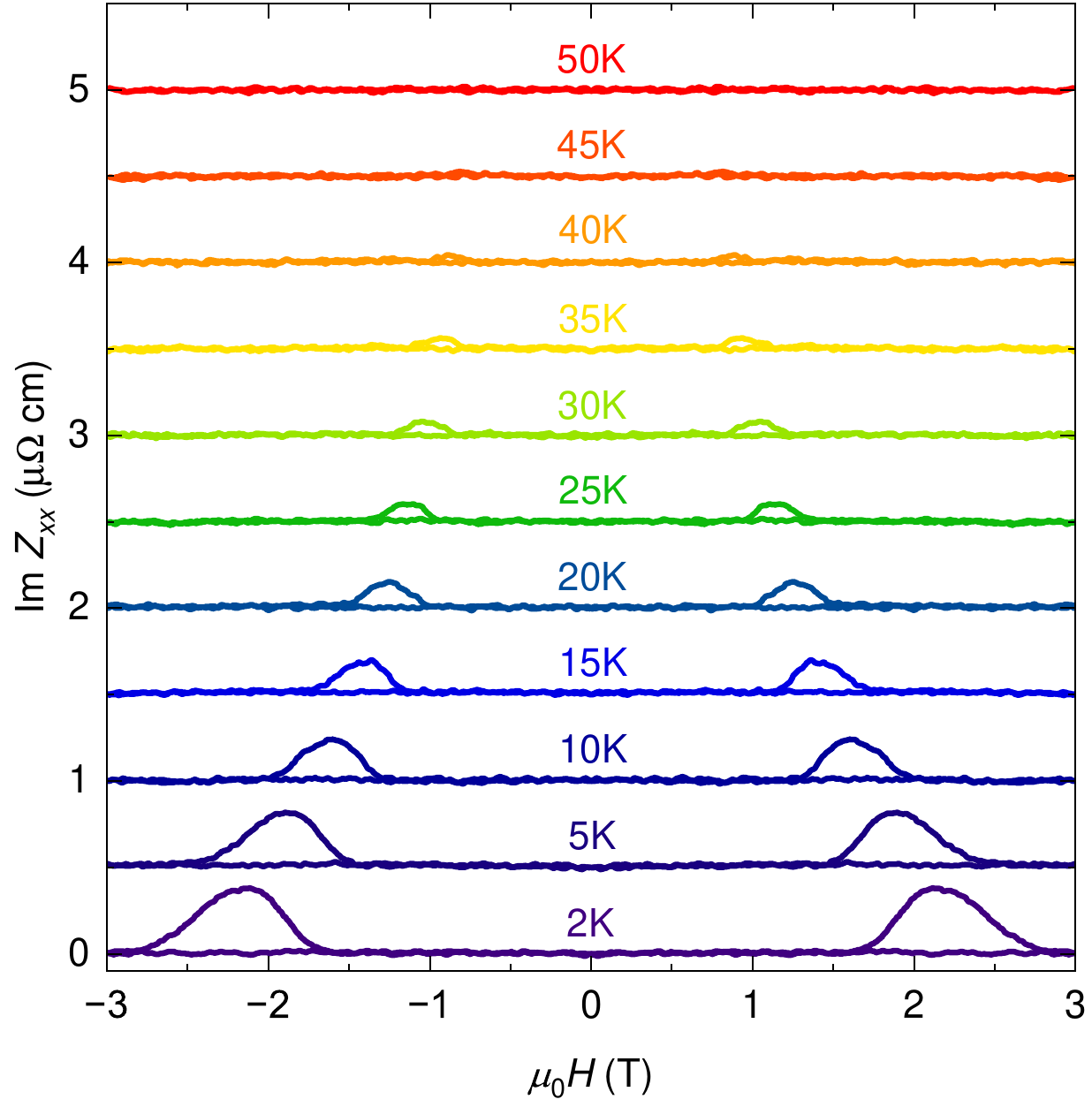}
\caption{\textbf{Temperature dependence of the imaginary part of the longitudinal impedance.} Magnetic field dependent imaginary impedance for serval different temperatures measured at \linebreak $j = 2.8\cdot10^7 \,\mathrm{Am}^{-2}$, $f = 10\,\mathrm{kHz}$, $d = 200$\,\textmu$\mathrm{m}$ and $A = 1.6$\,\textmu$\mathrm{m}^{2}$. The magnetic field $\mathbf{H}$ is along the out-of-plane direction.}\label{Supfig2}
\end{figure}

Figure~\ref{Supfig2} shows magnetic field dependent measurements of the complex impedance for different temperatures with constant current density of $j = 2.8\cdot10^7 \,\mathrm{Am}^{-2}$ and a frequency of $f = 10\,$kHz. For better visibility, the curves are displaced by $0.5$\,\textmu$\mathrm{\Omega~cm}$ in relation to each other. The inductive signal and the magnetic field value, where it reaches its maximum, $\mu_0H_\mathrm{max}$, both decline with increasing temperature. At $T = 50\,\mathrm{K}$, the inductive signal is barely detectable. The temperature dependence of $\mu_0H_\mathrm{max}$ can be used to calculate an effective temperature $T_\mathrm{eff}$ of the thin film in response to Joule heating, cf. \ref{sec: S1}. This temperature is shown in Fig.~\ref{Supfig_Iabh_2}~\textbf{b}. 

\begin{figure}[bh]
\centering
\includegraphics[width=0.533\textwidth]{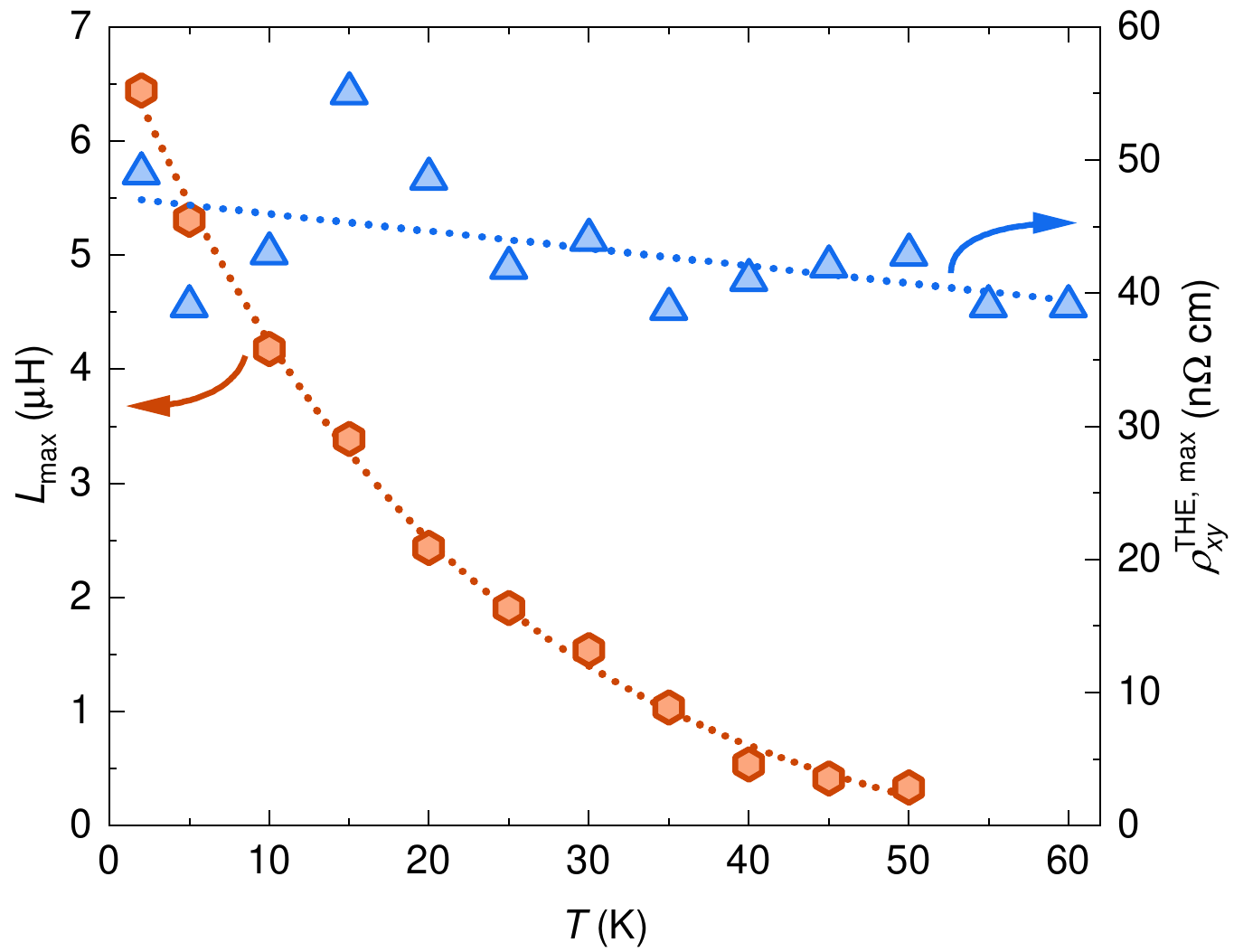}
\caption{\textbf{Induction and Hall effect signals of magnetic origin in SRO-SIO oxide bilayers.} Temperature dependent amplitude of the emergent electromagnetic induction signal (left axis), compared to the temperature dependence of the topological Hall effect (right axis). The peak values of the EEMI signal (orange) for several temperatures are taken from Main Text Fig~\ref{fig2}~\textbf{e}. }\label{Supfig6}
\end{figure}

Figure~\ref{Supfig6} shows the different temperature dependences of the measured inductance $L_{xx}^{\mathrm{max}}$ and the topological Hall-effect $\rho_{xy}^{\mathrm{THE, max}}$ respectively. The peak value of the THE decays slowly and quasi-linearly within this temperature range, while the EEMI signal drops rapidly with $T$. This indicates a drastic change in skyrmion dynamics, although the static properties of the spin texture change but little.
\clearpage
\subsection{Frequency dependence of the EEMI -- limitations of the experiment}\label{sec: fabh}
\begin{figure}[bh]
\centering
\includegraphics[width=0.6\textwidth]{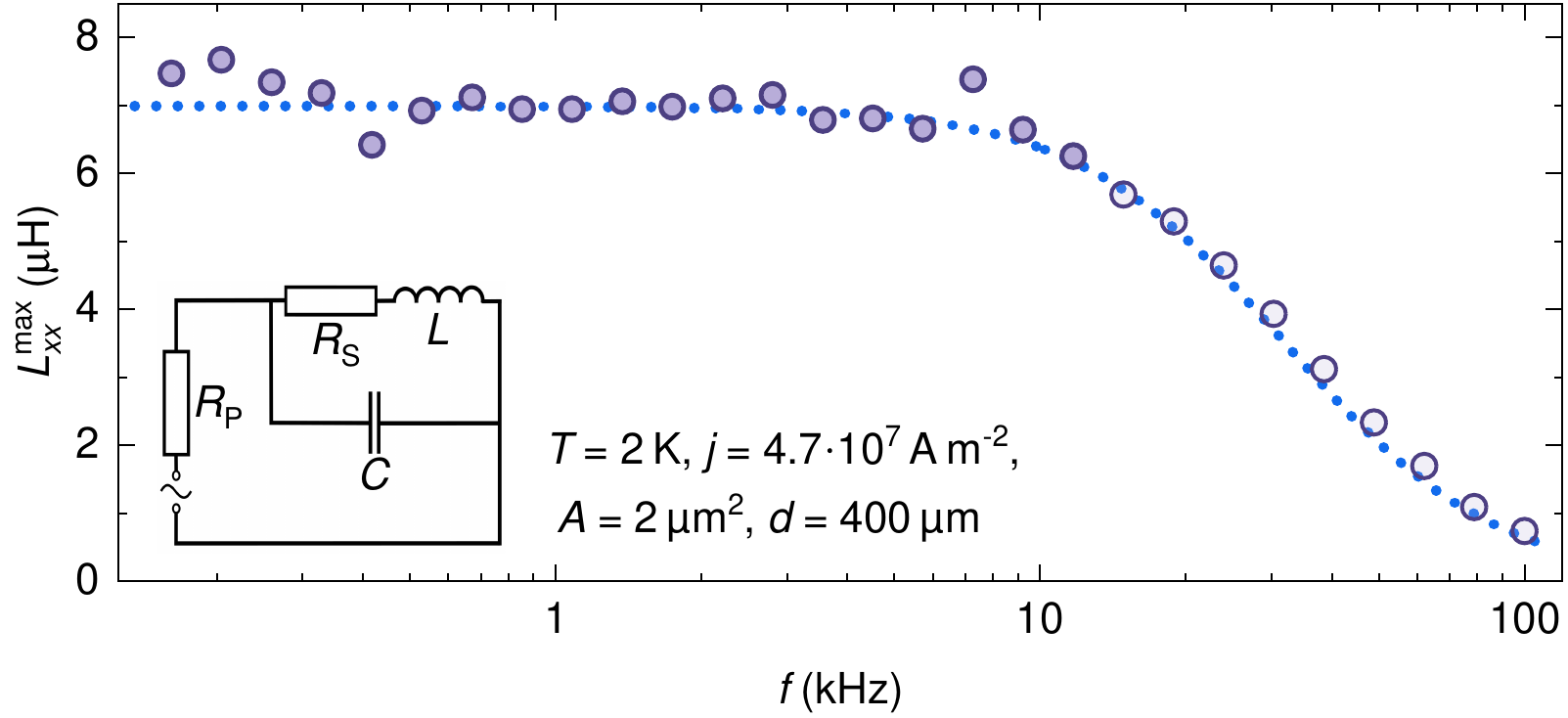}
\caption{\textbf{Frequency dependence of the EEMI.} Frequency dependence of the inductance $L^{\mathrm{max}}_{xx}$ measured at base temperature. The filled points show the expected frequency-constant induction behavior, whereas the empty points show the cut-off due to the parasitic capacitance. Inset: equivalent circuit diagram which approximates the observed frequency dependence (dotted blue line).}\label{Supfig_fabh}
\end{figure}

With a sample geometry of $A = 2$\,\textmu$\mathrm{m^2}$ and $d = 400$\,\textmu$\mathrm{m}$, the inductance $L^{\mathrm{max}}_{xx}$ at $T=2\,\mathrm{K}$ remains constant over nearly two orders of magnitude in frequency, shown in Fig.~\ref{Supfig_fabh}.
Capacitive couplings due to the wiring of the microstructure devices generates a low-pass filter with a cut-off frequency of approximately $f_{\mathrm{c}} = 20\,\mathrm{kHz}$, leading to a decrease in both the inductive signal and the resistivity at large $f$. The inset of Fig.~\ref{Supfig_fabh} shows the circuit used for simulations of the measured induction, where $R_{\mathrm{p}} = 50\,\mathrm{k\Omega}$ denotes the shunt resistance, while $R_{\mathrm{S}}$ and $L_{\mathrm{S}}$ represent the sample resistance and the measured emergent inductance, respectively. The parallel connected capacitance has a value of $C \approx 4.5\,\mathrm{nF}$, which is realistic for our experimental setup.
\clearpage
\subsection{Background subtraction for complex impedance data}\label{sec: S3}
In our measurement circuit, the experimental impedance of interest arises in the bilayer, but parasitic impedances are also present. In addition to contact resistances and the wiring of the cryostat, the microstructured sample itself creates a capacitive contribution, which is connected in parallel to the main circuit. Therefore, a temperature and frequency dependent, but magnetic field-independent background is subtracted from the total impedance to obtain the impedance of interest in each measurement. Additionally, we adjusted for each individual measurement the phase relation $e^{\imath \phi}$ of the injected current and the voltage response in the high field ferromagnetic saturated state, to ensure that Im$Z$ is minimized, in order to be able to measure small changes with maximal sensitivity. The measured signal has the form
\begin{align*}
    \Tilde{Z} = e^{\imath \phi} \cdot \left( \mathrm{Re} Z^{*} + \imath \mathrm{Im}Z^{*} \right),
\end{align*}
where $\Tilde{Z}$ is the impedance detected by the Lock-in amplifier and $Z^{*}$ is the circuit impedance including all parasitic contributions. The phase shift $\phi$ from the lockin amplifier leads to a mixing of the Re$Z^{*}$ and Im$Z^{*}$ parts of the circuit impedance. We compensated this effect by a simple phase reversal as follows:
\begin{align*}
    \mathrm{Re} Z^{*} = \mathrm{Re} \Tilde{Z} \cdot \cos \phi - \mathrm{Im} \Tilde{Z} \cdot \sin \phi \\
    \mathrm{Im} Z^{*} = \mathrm{Im} \Tilde{Z} \cdot \cos \phi - \mathrm{Re} \Tilde{Z} \cdot \sin \phi
\end{align*}
Since the parasitic contributions to the imaginary impedance do not strongly depend on the magnetic field, they are assumed to be constant and subtracted from $\mathrm{Im}Z^{*}$, allowing the desired inductive contribution to be isolated for the given temperature and frequency.
\clearpage
\subsection{Skyrmion motion in the pinned regime}\label{sec: sk_motion}
\begin{figure}[b]
\centering
\includegraphics[width=0.7\textwidth]{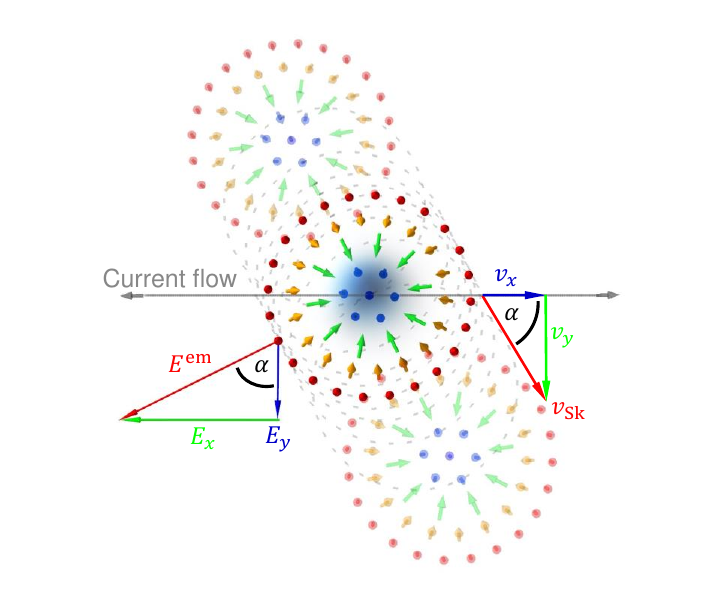}
\caption{\textbf{Bird's-eye view of the skyrmion dynamics in the pinned regime for SRO-SIO oxide bilayers.} As described in \ref{sec: S3}, the experiment indicates a skyrmion motion with velocity $\bm{v}_{\mathrm{Sk}}$, which is not perfectly perpendicular to the applied electric current $j$; the angle $\alpha$ is estimated to be $\sim 80^\circ$, which provides two vector components of the resulting emergent electric field $\bm{E}^{\mathrm{em}}$.  }.\label{EEMI_calc}
\end{figure}
The skyrmion motion in the pinned regime is decisive for the observed emergent inductance in our SRO-SIO oxide bilayers. Considering the out-of-plane emergent magnetic field $\mathbf{B}^{\mathrm{em}} = B^{\mathrm{em}}_z\cdot\mathbf{e}_z$, the emergent electric field $\mathbf{E}^{\mathrm{em}}$ depends on the skyrmion velocity $\mathbf{v}_{\mathrm{Sk}}$ as described by Eq.~(\ref{eq:vsk}) of the Main Text. Figure \ref{EEMI_calc} is a schematic bird's-eye view of the oscillating skyrmion motion in the pinned regime, and the resulting emergent electric fields. Here we assume a deviation of $\alpha$ from $90^\circ$, i.e. a low-symmetry condition, where the skyrmion motion in the pinned regime is not perfectly perpendicular to the applied electric current density $\mathbf{I} = j_x/A\cdot\mathbf{e}_x$. The transverse skyrmion velocity $v_y$ in Eq.~(\ref{eq:vsk}) of the Main Text accounts for the longitudinal emergent electric field $E^{\mathrm{em}}_{x}$; meanwhile, $-v_x$ accounts for $E^{\mathrm{em}}_{y}$. We calculate the emergent electric field from $E_{j} = I_i\cdot\mathrm{Im}Z_{ij}\cdot A/d$ and the emergent magnetic field $B^{\mathrm{em}}$ from the experimentally observed THE. Thus, we estimate that -- in the pinned regime, for our SRO-SIO oxide bilayers -- the skyrmions move at an average speed of $v_{\mathrm{sk}} = 0. 025\,\mathrm{m/s}$ and at an angle of about $\alpha = \arctan\left(E^{\mathrm{em}}_{x}/E^{\mathrm{em}}_{y}\right)\approx 80^{\circ}$ with respect to the applied current -- raw data for the calculation are given in Fig.~\ref{fig2}~\textbf{a} and \textbf{b} of the Main Text.

\clearpage
\subsection{General Information}\label{sec: S4}

\begin{table}[th]
    \centering
    \caption{\textbf{Samples and microstructures used in this study.}}
    \label{tabS1}
    \begin{tabular}{c|c|c}
        Sample number &  $A\,\left( \text{\textmu}\mathrm{m^2}\right)$& Electrode distances $d\left( \mathrm{\text{\textmu} m}\right)$\\
        \hline
              & 1.0 & 50, 200, 300, 450 \\
        \#1  & 1.2 & 300 \\                        
              & 1.5 & 250, 500 \\
        \hline
              & 0.5 & 50, 200, 300, 450 \\
        \#2   & 1.0 & 50, 200, 300, 450 \\          
              & 2.0 & 50, 200, 300, 450 \\
        \hline
        \#3   & 0.5 & 50, 200, 300, 450 \\         
              & 1.0 & 50, 200, 300, 450 \\  
        \hline
              & 1.0 & 100, 200, 300, 400, 500 \\
              & 1.2 & 100, 200, 300, 400, 500 \\
        \#4   & 1.6 & 100, 200, 300, 400, 500 \\    
              & 2.0 & 100, 200, 300, 400, 500 \\
              & 3.0 & 300\\
        \hline
        \#5   & 1.0 & 100, 200, 300, 400, 500 \\    
    \end{tabular}
\end{table}

\begin{figure}[th]
\centering
\includegraphics[width=0.6\textwidth]{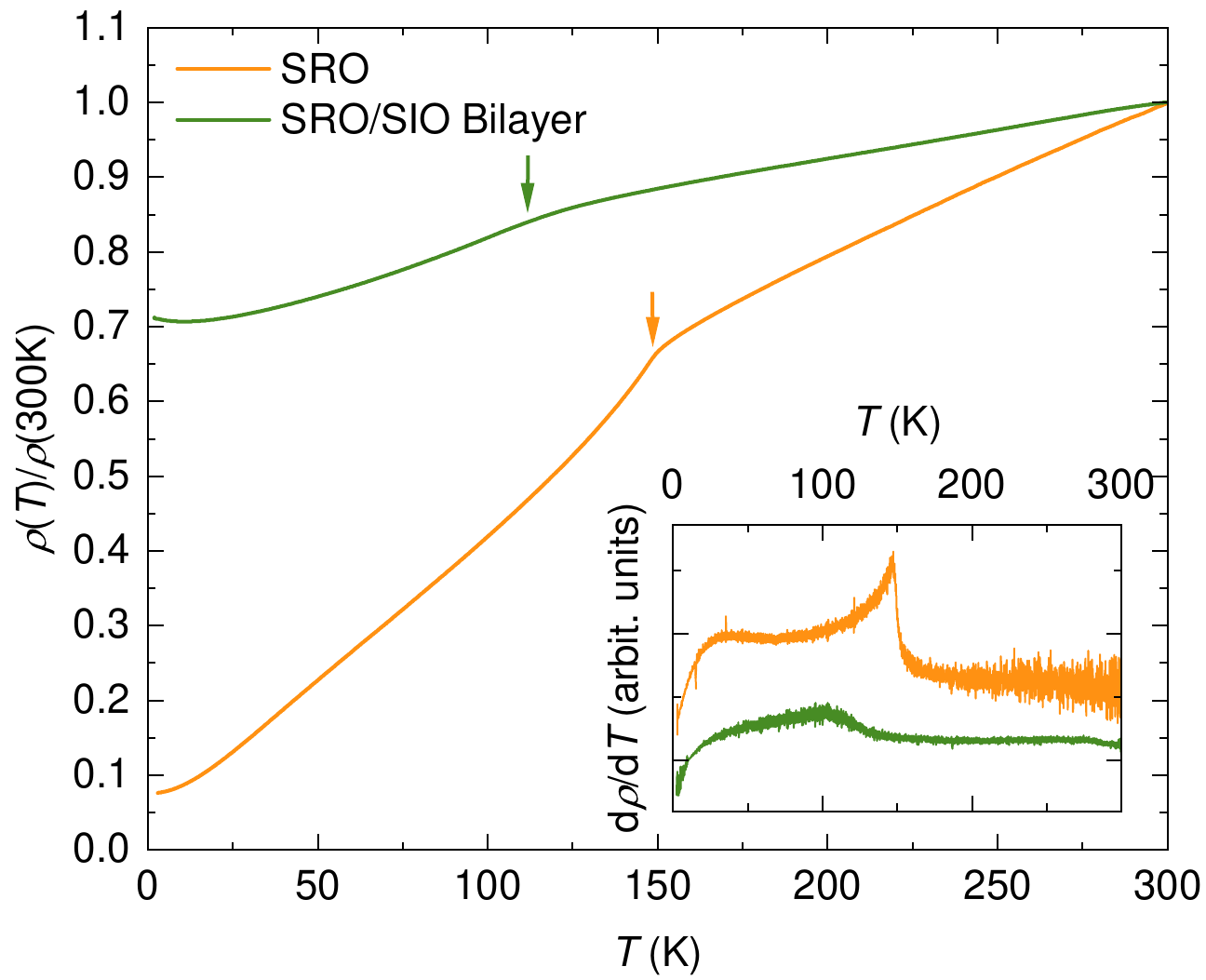}
\caption{\textbf{Temperature dependent electrical resistivity measurements for SRO thin films and for SRO-SIO oxide bilayer devices.} Resistivity curves of a $22\,\mathrm{nm}$ SRO thin film (orange line) and SIO-SRO sample \#4 of this study (green line), cf. Table~\ref{tabS1}. The traces were scaled to the same value at room temperature. The pure SRO sample shows a magnetic phase transition at about $T\sim150\,$K, marked by a sharp kink, cf. derivative plot in the inset. Meanwhile, the phase transition in the SRO-SIO bilayer, at around $T\sim110\,$K, is broader. This is due to the reduced thickness of the active layer in the bilayer device.}\label{Supfig4}
\end{figure}

\begin{figure}[bh]
    \centering
    \includegraphics[width=\textwidth]{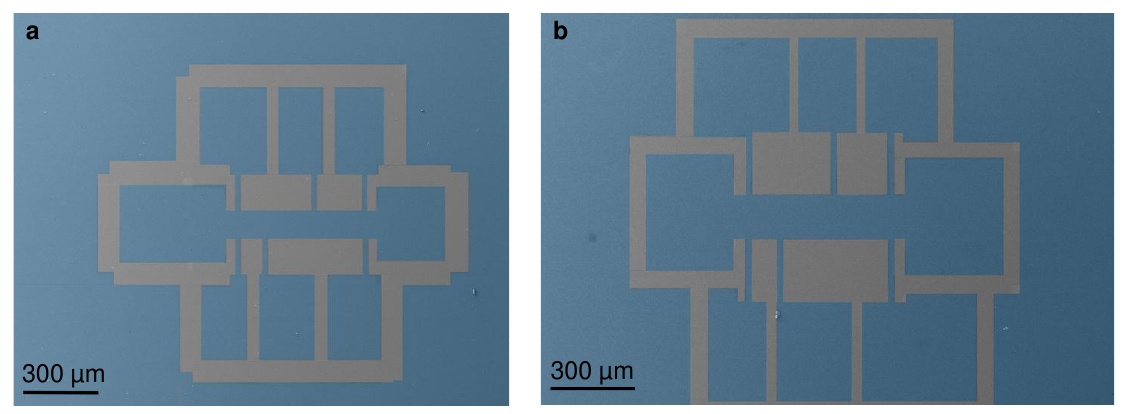}
    \caption{\textbf{Scanning electron microscopy (SEM) images of the microstructured samples.} Representative SEM images of two different microstructures: Sample \#5 with $A = 1.0$\,\textmu$\mathrm{m}^2$ (\textbf{a}) and sample \#4 with $A = 1.6$\,\textmu$\mathrm{m}^2$ (\textbf{b}).}
    \label{Supfig5}
\end{figure}

\begin{figure}[bh]
    \centering
    \includegraphics[width=\textwidth]{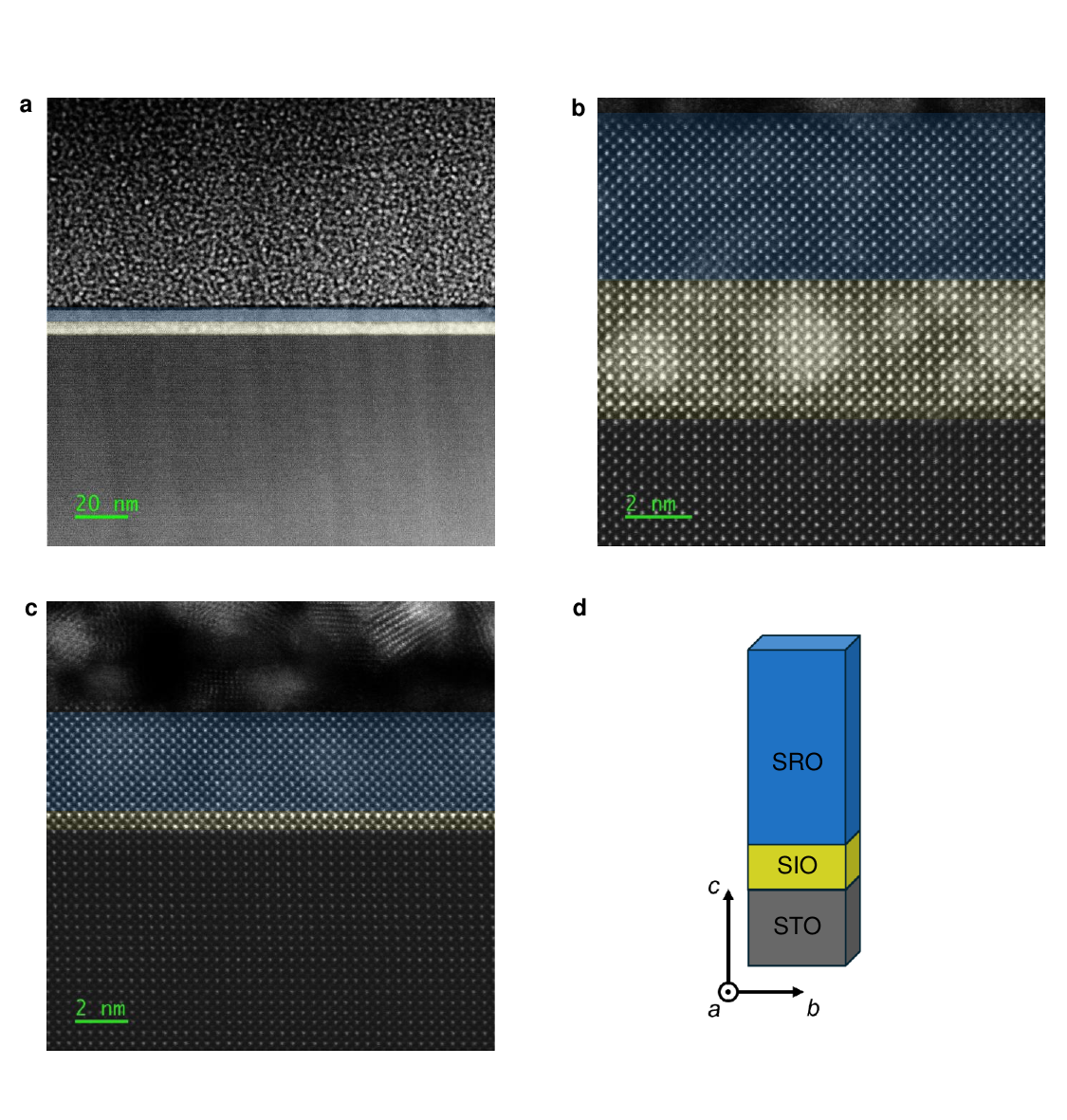}
    \caption{\textbf{Additional TEM-images of SRO-SIO-bilayer structures.} \textbf{a}, \textbf{b} Sample \#4 with a structure composed of $10$\,uc SIO and $10\,$uc SRO, with high structural quality over a large area (a) as well as in a close-up (b). \textbf{c} Sample \#1 with a $2\,$uc SIO and $10\,$uc SRO structure. \textbf{d} Schematic bilayer geometry with crystal axes and explanation of false color code used for visualizing the TEM images.}
    \label{Supfig3}
\end{figure}
\end{document}